\documentclass{webofc}

\usepackage[varg]{txfonts}   
\usepackage{hyperref}
\usepackage{url}
\usepackage{listings}
\usepackage{xcolor}

\makeatletter
\g@addto@macro\UrlBreaks{\do\-}  
\makeatother

\hypersetup{colorlinks=true,citecolor=blue,urlcolor=blue,linkcolor=blue}
\RequirePackage{orcidlink}
\definecolor{SkoleBrown}{RGB}{150,100,50}
\definecolor{SkoleYellow}{RGB}{232,183,3}
\definecolor{SkoleOrange}{RGB}{212,80,0}
\colorlet{SkoleLightBrown}{SkoleBrown!50!SkoleYellow}
\definecolor{lightgrey}{cmyk}{0,0,0,0.5}
\colorlet{codeblockcolorbase}{SkoleLightBrown!80!black}
\colorlet{codeblockcolor}{codeblockcolorbase!02}

\begin{document}
\title{GPU Acceleration of Awkward Arrays: Using Python cuda.compute}
%
%

\author{\firstname{Maksym} \lastname{Naumchyk}\inst{1}\orcidlink{0009-0000-3348-859X}\fnsep\thanks{\email{maxymnaumchyk@gmail.com}} \and
        \firstname{Ianna} \lastname{Osborne}\inst{1}\orcidlink{0000-0002-6955-1033}\fnsep\thanks{\email{Ianna.Osborne@cern.ch}}
}

\institute{Princeton University 
          }

\abstract{Awkward Array is a widely used library in high-energy physics (HEP) for representing and manipulating nested, variable-length data in Python. Previous CHEP contributions have explored GPU acceleration for Awkward Array, demonstrating the feasibility and performance benefits of CUDA-based backend while also identifying limitations related to irregular data access, fine-grained kernel launches, and composability of operations. In this contribution, we present recent developments that build directly on these earlier efforts by introducing a CUDA execution model for Awkward Array based on the Python CUDA Core Compute Libraries (CCCL).

Using CCCL, we eliminate the need for custom CUDA kernels and can instead use a high-level Python interface. The CCCL-based approach also enables fusion of multiple Awkward operations into a reduced number of CUDA kernels, addressing kernel launch overhead observed in earlier GPU implementations. Lazy execution allows expression graphs to be constructed and optimized prior to kernel generation, improving performance for analysis workflows involving jagged arrays, combinatorial operations, and reductions. In contrast to earlier approaches, this design also emphasizes extensibility, allowing user-defined Python code to be incorporated into GPU execution paths with minimal boilerplate and without breaking existing analysis semantics.

We present performance studies that demonstrate improvements over previously reported eager GPU execution strategies for representative HEP analysis patterns. These developments extend the GPU capabilities of Awkward Array toward a more composable and sustainable backend, aligned with the needs of Python-based analysis at the HL-LHC and beyond.

We thank the NVIDIA team for support and collaboration in integrating CCCL into Awkward, developing CUDA kernels and providing guidance on GPU optimization strategies. Their contributions are gratefully acknowledged.
}
\maketitle

\section{Prologue}
\label{prologue}
Each of the different Awkward functions (\textit{sum, mean, etc.}) consists of one or multiple \textit{kernels}. Historically, these kernels were hand-written directly in CUDA C++. However, this implementation had its own drawbacks: raw CUDA is hard to write, hard to optimise, and hard to maintain. So the idea was to replace the existing kernels with cuda.compute --- a Python interface to CUDA’s parallel algorithms. This approach would greatly reduce the codebase and make it much easier to maintain and develop new kernels. It would even bring performance improvements coming from using optimised cuda.compute algorithms and reducing kernel launches. The idea took shape at the ACAT workshop~\cite{acat}, where the limitations of the existing approach became the starting point for this work.

The next two sections introduce Awkward itself and CUDA kernels in more detail. If you are already familiar with these concepts, jump straight to Section~\ref{cuda.compute}.

\section{Introduction or is your data feeling awkward?}
\label{intro}
\begin{figure}
    \centering
    \includegraphics[width=1\linewidth]{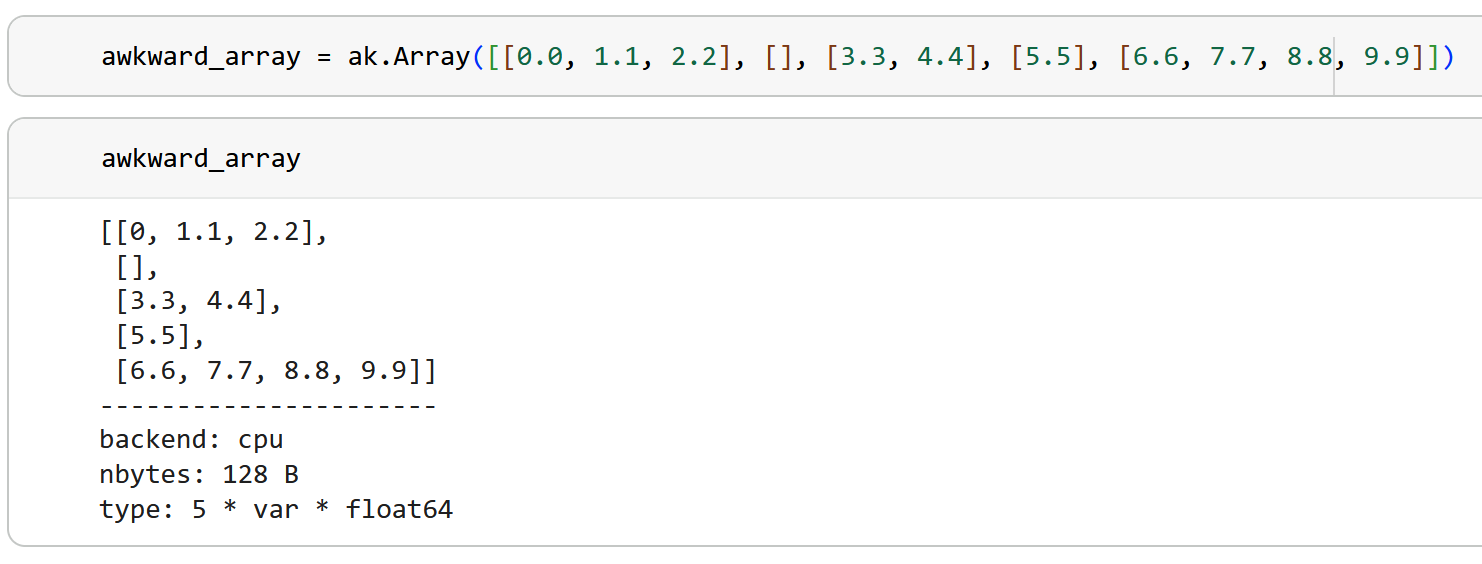}
    \caption{Using non-rectangular arrays with Awkward}
    \label{ak}
\end{figure}

Let us start from afar. Numbers. You have data you want to analyse. One of the most common tools to do that now is the NumPy library\cite{numpy}. It lets you structure your data into arrays. For example, you can have an array like that:
\begin{lstlisting}
numpy_array = np.array([0.0, 1.1, 2.2, 3.3, 4.4, 5.5, 6.6, 7.7, 8.8, 9.9])
\end{lstlisting}
Real-world data is not always so simple. Take this sentence, for example: five words, each with three to eight letters. If we want to convert this information into arrays of numbers, we can get something like this:
\begin{lstlisting}
[
  ['T', 'a', 'k', 'e'],
  ['t', 'h', 'i', 's'],
  ['s', 'e', 'n', 't', 'e', 'n', 'c', 'e'],
  ['f', 'o', 'r'],
  ['e', 'x', 'a', 'm', 'p', 'l', 'e']
]
\end{lstlisting}
Now, if we try using NumPy to store this sentence, we will run into an error:
\begin{lstlisting}
ValueError: The requested array has an inhomogeneous shape
\end{lstlisting}
What does it mean? \textit{Inhomogeneous shape}. It means that each of our subarrays (words) has a different length. Such arrays are sometimes called non-rectangular. Some libraries call them jagged or ragged arrays~\cite{ragged}. We call them Awkward Arrays~\cite{awkward}, and we have a library with the same name where you can use them! (See Figure~\ref{ak}.)

\begin{figure}
    \centering
    \includegraphics[width=1\linewidth]{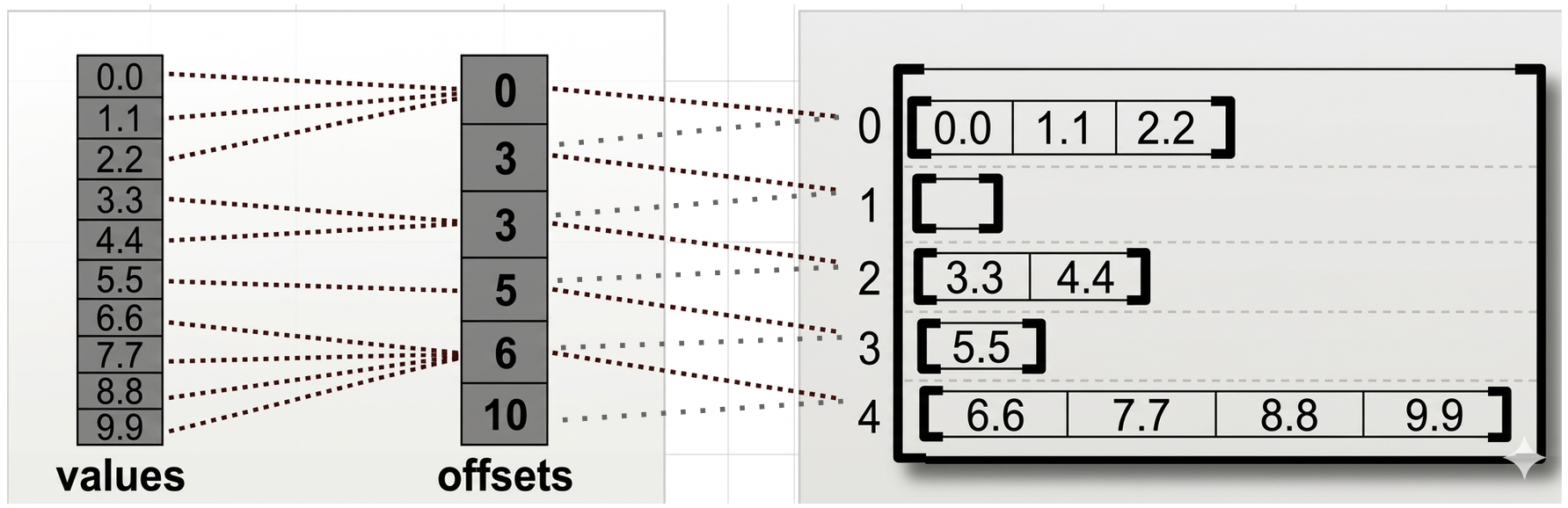}
    \caption{Inner structure of Awkward arrays}
    \label{ak_structure}
\end{figure}

Let us take a more detailed look at how the Awkward arrays work from the inside. Internally, they are built from the same NumPy arrays. We flatten all of our data into a single array. This is marked as "values" in Figure~\ref{ak_structure}.

We also create one new NumPy array that stores information about the "raggedness" of the data — basically, the beginning and end index of each of our subarrays. We call them "\textit{offsets}". So, from Figure~\ref{ak_structure}: the first subarray has a start offset of "0" and an end offset of "3". The second array has both a start and end offset of "3", meaning that it is an empty array, and so on. As you can see, each subarray has its own pair of offsets. Now, an Awkward user does not need to know about such specifics. Why are we mentioning this here then? It will come in handy later on in this article.

\begin{figure}
    \centering
    \includegraphics[width=1\linewidth]{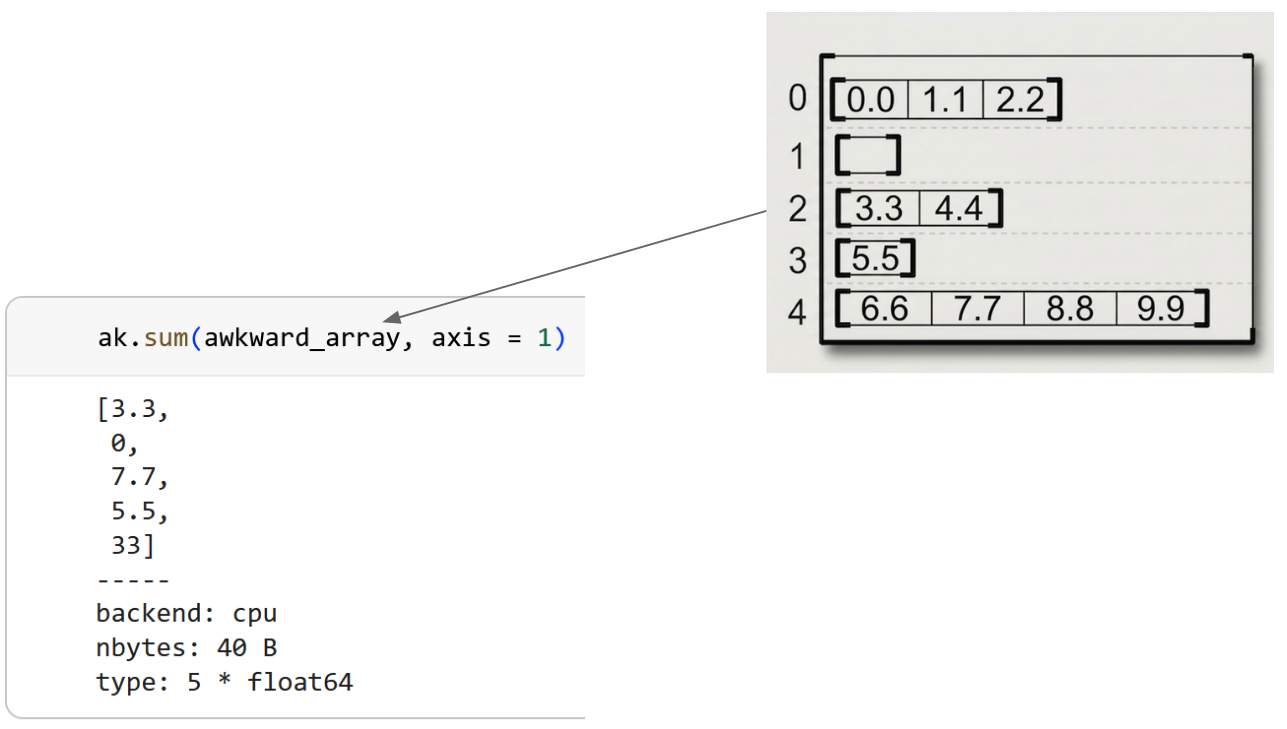}
    \caption{Finding a sum on axis 1 (for each subarray)}
    \label{sum}
\end{figure}

We loaded our data. Now we finally want to do something with it. Awkward has a huge variety of NumPy-like operations. You can calculate combinations, mean, argmax of arrays, and many more. Let us take a simple summation as an example. Using it is as easy as shown in Figure~\ref{sum}.

As a result, we get the number "3" for the sum of all elements of the first subarray (0+1+2), zero for an empty array, and so on. Now, once again, let us look deeper into how the summation operation works inside Awkward.

\section{CUDA kernels}
\label{sec-1}
\begin{figure}
    \centering
    \includegraphics[width=1\linewidth]{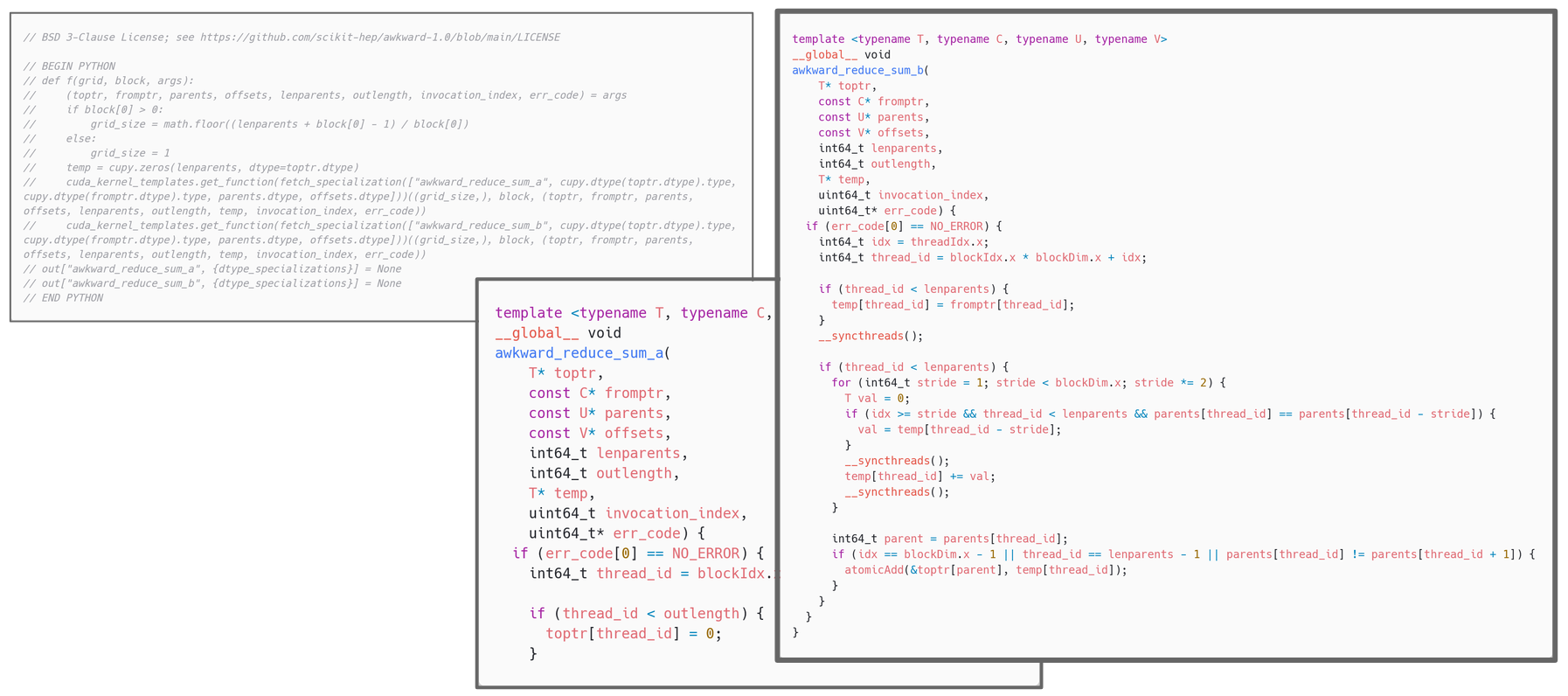}
    \caption{The whole ak.sum() operation using raw CUDA\cite{CUDA} kernels}
    \label{cuda}
\end{figure}

Figure~\ref{cuda} shows how Awkward handled a summation operation just a while back.

You usually have a couple of C++ functions that parallelise their specific tasks. Here we have two such functions: \textit{awkward\_reduce\_sum\_a} and \textit{awkward\_reduce\_sum\_b}. Such functions are also called \textit{kernels}. Almost all of the code in these kernels exists just to set things up. Computing thread positions, tracking subarray boundaries, and keeping threads in sync. Only one line at the end does the real accumulating work:
\begin{lstlisting}
atomicAdd(&toptr[parent], temp[thread_id]);
\end{lstlisting}
This line adds each thread's partial sum into the correct output slot.

Using raw CUDA is cumbersome. You need to be an expert to write CUDA code. You need to constantly think about the optimisations that could take place. Maintaining and updating such code is also time-consuming. A new CUDA package can help us with that.

\section{The cuda.compute package}
\label{cuda.compute}
\begin{figure}
    \centering
    \includegraphics[width=1\linewidth]{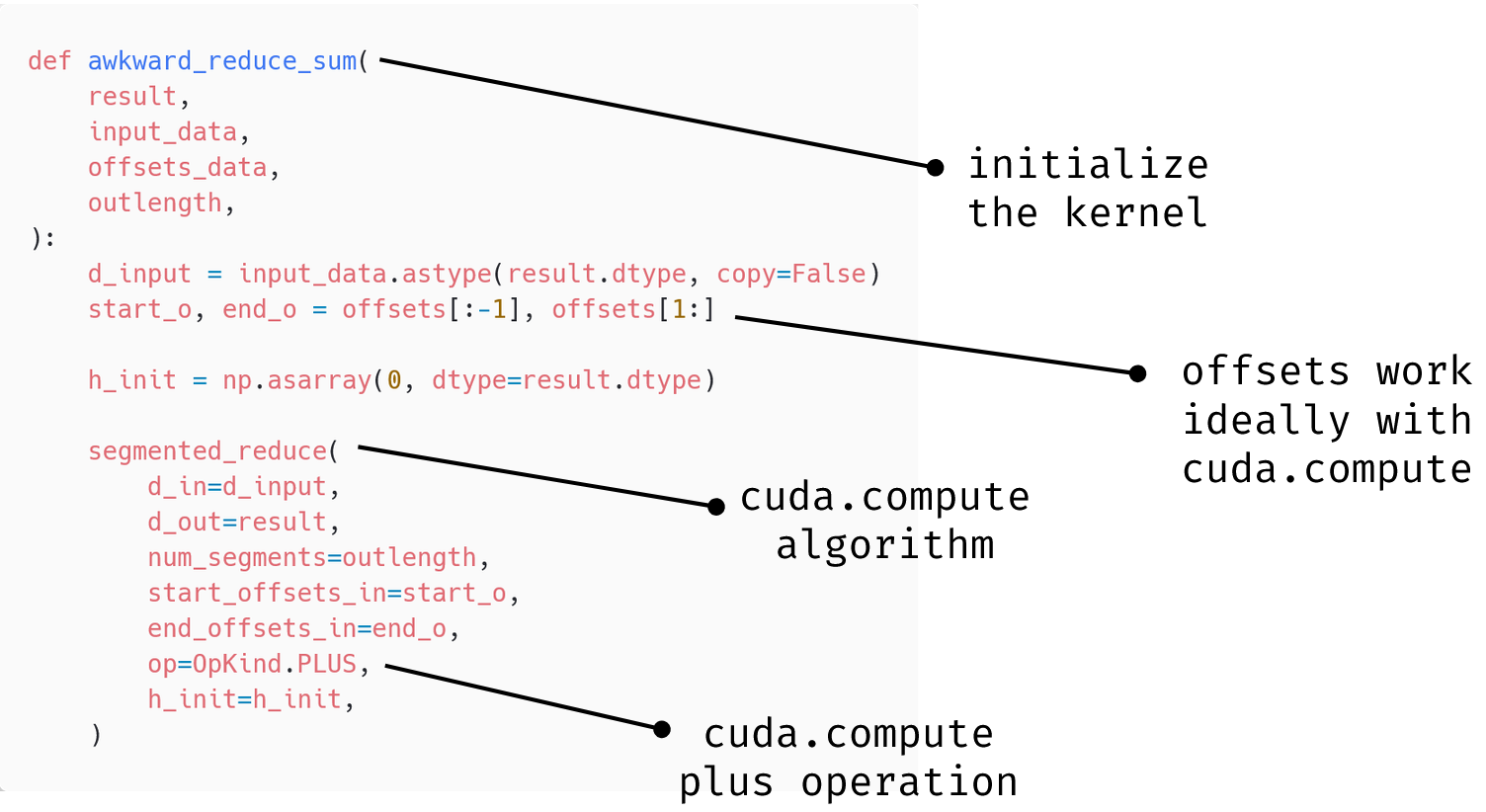}
    \caption{Implementation of a summation operation using cuda.compute}
    \label{reduce_sum}
\end{figure}

Recently, NVIDIA has been developing a new package called cuda.compute. It provides a simple Python interface to write CUDA kernels. Let us take our previous summation operation and rewrite it using cuda.compute (Figure~\ref{reduce_sum}).

The code itself is short: we pass in the values array together with the start and end offsets (Section~\ref{intro}), an initial value \textit{h\_init} for the reduction (summation in our case), and call \textit{segmented\_reduce}. This is one of the algorithms cuda.compute provides.

This maps so well onto Awkward arrays because of their structure. Recall from Section~\ref{intro} that an Awkward array is really just a flat "values" array and a set of offsets, marking where each subarray begins and ends. A \textit{segmented reduction} is built to do exactly this. Given a flat array and a set of segment boundaries, it reduces each segment independently and in parallel. The "segments" that cuda.compute expects are precisely the subarrays that Awkward's offsets already describe, so no restructuring or custom indexing logic is needed.

This gives us two practical advantages. First, the whole reduction runs as a single kernel launch, rather than the several kernels that our old raw CUDA implementation needed. Second, the same \textit{segmented\_reduce} call can be reused for other reduction-like operations (\textit{mean, min, max, etc.}). Just by changing the operator, instead of hand-writing a new kernel for each one.

With that, we successfully rewrote the Awkward summation operation. Outside, we only use Python and cuda.compute algorithms. Inside, these algorithms use highly optimised C++ libraries: CUB and Thrust. This way we leave the development of new algorithms and all the optimisations to NVIDIA developers and concentrate only on the correct kernel behaviours.

\section{Speed comparison}
\begin{figure}
    \centering
    \includegraphics[width=1\linewidth]{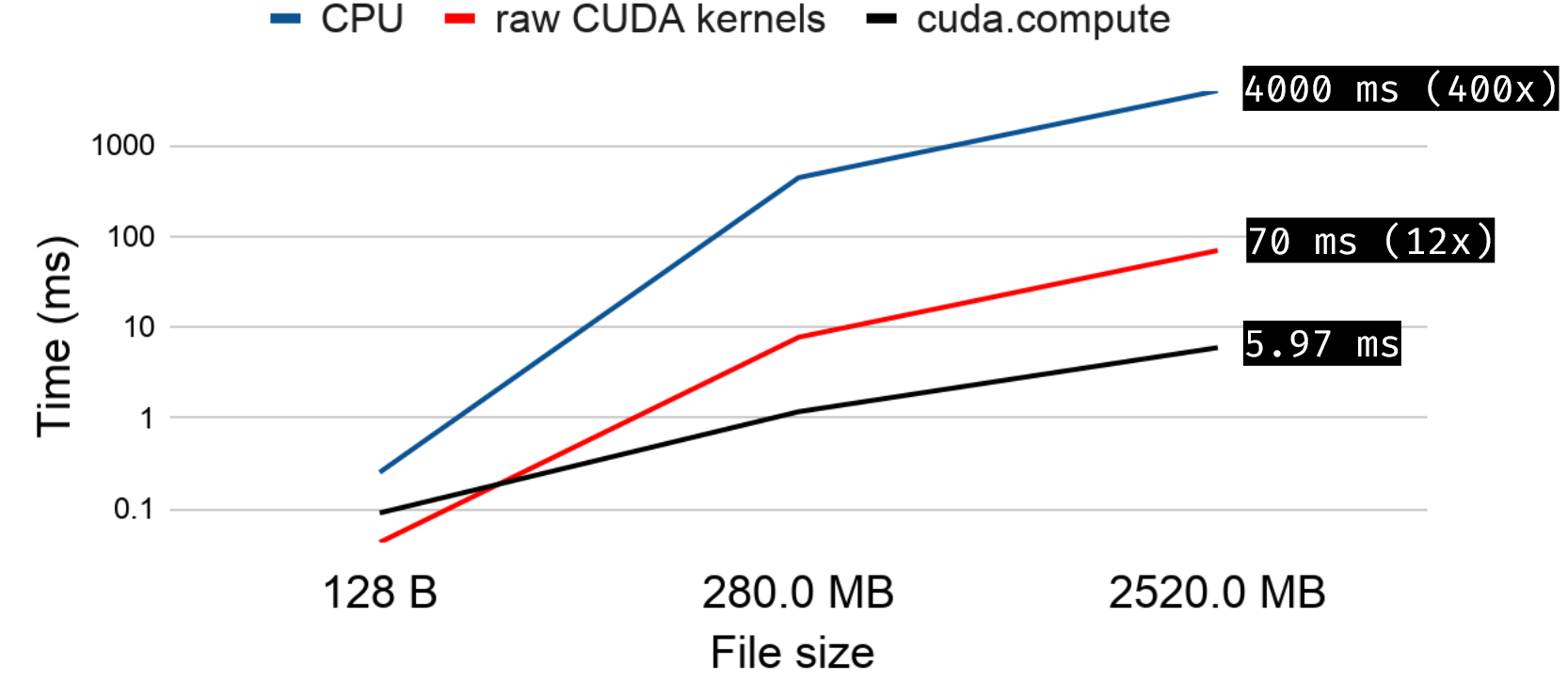}
    \caption{Benchmarking the speed of ak.argmax() using arrays of different sizes}
    \label{argmax}
\end{figure}

In this section, let us explore how the cuda.compute-based implementation performs compared to the old raw CUDA kernels. Figure~\ref{argmax} shows the use of another Awkward operation (argmax) for arrays of different sizes.

In terms of performance, the cuda.compute-based implementation is around 12 times faster than the raw CUDA kernels it replaces, and around 400 times faster than the equivalent implementation on a CPU. The bigger the arrays, the better the performance gain we get. This benchmark \cite{argmax-bench} was run on a personal computer with an Intel Core i7-12700 CPU and an RTX 3080 GPU. \par
\begin{figure}
    \centering
    \includegraphics[width=1\linewidth]{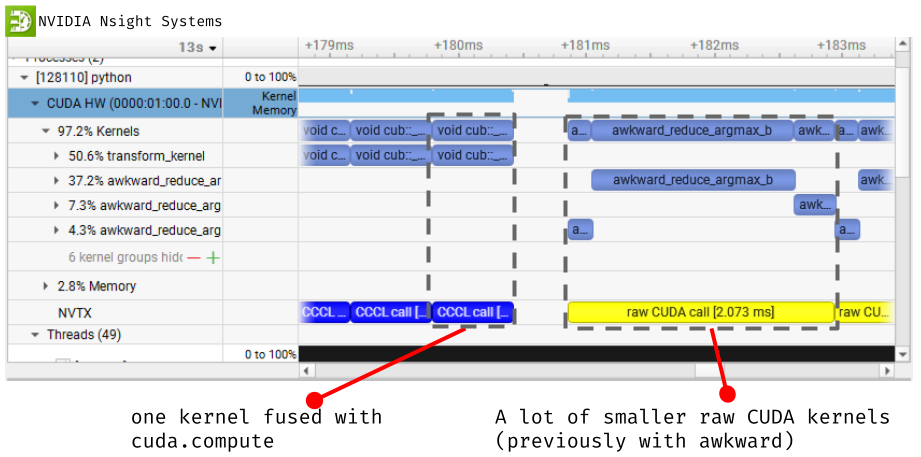}
    \caption{Benchmarking the ak.argmax() operation using NVIDIA Nsight Systems}
    \label{nsight}
\end{figure}

One of the biggest reasons for 12 times improvement is "kernel fusion". Figure~\ref{nsight} shows that the old implementation of the argmax operation needs three kernels, and the new cuda.compute implementation uses just one kernel (the same \textit{segmented\_reduce} we saw before). 

Every kernel launch has a fixed cost, but the bigger issue is memory. Each kernel has to write its results back to the GPU's global memory before the next kernel can read them in. Memory access is typically the slowest part of any GPU operation. With three separate kernels, this happens twice. Once between each pair of kernels. Fusing all three steps into a single \textit{segmented\_reduce} call removes both of these round trips, keeping intermediate values on-chip instead. Generally, when working with CUDA, the fewer kernels, the better. \par
On top of that, \textit{segmented\_reduce} itself benefits from years of tuning inside CUB and Thrust. Careful use of shared memory, coalesced memory access and load balancing across thread blocks are built into cuda.compute.

\begin{figure}
    \centering
    \includegraphics[width=1\linewidth]{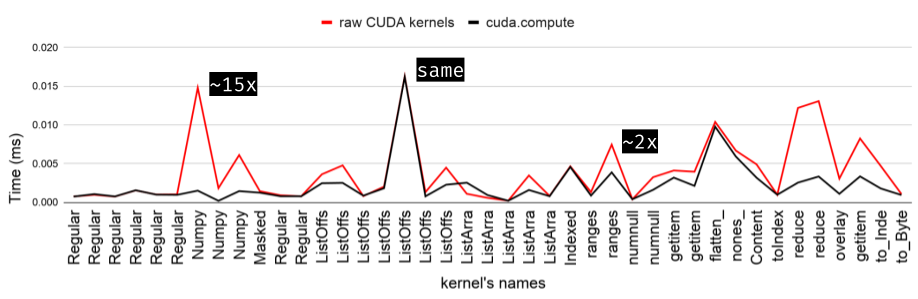}
    \caption{Comparing the speed of ~40 different kernels.}
    \label{bench}
\end{figure}

Let us perform some more benchmarks. We will take around 40 kernels and compare their speeds for two different implementations. For the tests, we will be using big sample arrays (several GBs). Figure~\ref{bench} shows an improvement in speed of up to 15 times. The cause of this is the same as described above. Fewer kernel launches and better-optimised memory access. For more benchmarks using actual physics analysis, see the work of Kelci Mohrman et al. \cite{sci-bench} or the SciPy blog from Ianna Osborne and Ashwin Srinath \cite{cuda-compute-blog}.

\section{Kernel development process}
\label{kernel-development-process}

At the time of writing this article, around 90\% of all the Awkward kernels are already migrated. For the remaining kernels, using raw CUDA is still beneficial.
\begin{lstlisting}
cupy.RawKernel()
\end{lstlisting}
The most prominent examples of such cases are the kernels that raise errors. cuda.compute's algorithms have no built-in way for a thread to report a failure back to the caller. Awkward's raw CUDA kernels handle this with a shared error flag instead.

It is also worth noting that, in some cases, using CuPy alone can be beneficial.
\begin{lstlisting}
def awkward_ListArray_rpad_and_clip_length_axis1(
    tomin, fromstarts, fromstops, target, lenstarts
):
    counts = fromstops[:lenstarts] - fromstarts[:lenstarts]
    tomin[:1] = cp.sum(cp.maximum(counts, target)).astype(tomin.dtype, copy=False)
\end{lstlisting}
This is just an example of one of the kernels that Awkward uses. The exact implementation details are not important here. Just note that different cupy operations are used here, such as \textit{cp.sum()}, \textit{cp.maximum()}, and \textit{slicing}.\par
The cuda.compute package is still new and is currently being actively developed. That is why some workarounds like these can take place now. We (the Awkward library team) are working very closely with the NVIDIA team to address any bugs or performance issues we encounter. Speaking of which, CUDA Python (a parent package of cuda.compute) just had its first release \cite{blog} (at the time of writing this article). 

\section{Next steps – improving performance}
In this section, we will explore how to continue improving the GPU performance of Awkward Arrays.
\begin{itemize}
\item First, we will extensively benchmark the implementations of new kernels. Validation by Awkward users is also needed to make sure no calculations are affected by the changes. Based on these results, we can identify the existing bottlenecks and work on them.
\item Right now, all the cuda.compute kernels are JIT (just-in-time) compiled. We will be integrating ahead-of-time compilation when this gets officially supported.
\item In the example that was used in this article (an implementation of summation), we used only one \textit{segmented\_reduce} operation. More are required for some of the more complex Awkward functions. Going one step further, we can fuse even the cuda.compute operations themselves — for example, using Parrot \cite{parrot}, another NVIDIA library. Since it is still in early development, other solutions are possible, such as lazy execution built in directly inside Awkward.
\end{itemize}

\section{Acknowledgments}
This work was supported by the National Science Foundation under Cooperative Agreement PHY-2323298 (IRIS-HEP).\par
On behalf of the Awkward team, we thank the NVIDIA team (Jake Hemstad, Ashwin Srinath, et al.) for their help and support with any obstacles we encountered during this project.\par
Thanks to my mentor, Ianna Osborne, for helping and advising me on this project. Thanks to the IRIS-HEP grant for sponsoring me on this journey, giving me an opportunity to work with amazing people of the HEP community and to present the results of my work at the 28th Conference on Computing in High Energy and Nuclear Physics (CHEP 2026) in Bangkok.

%
%

\end{document}